\documentclass[aps,prl,twocolumn,nofootinbib,showkeys]{revtex4}
\usepackage{epsfig}
\usepackage{graphicx}
\usepackage{amssymb}
\usepackage{color}
\usepackage{amsmath}
\usepackage[colorlinks=true]{hyperref}
\usepackage{graphicx}
\usepackage{afterpage}
\usepackage{bm}
\usepackage{xspace}

\newcommand{\nuc}{$^{178}$Os\xspace}

\graphicspath{{figures/}}

\begin{document}

\author{Rapha\"el-David Lasseri$^*$} \email{raphael.lasseri@cea.fr}
\affiliation{ESNT, CEA, IRFU, D\'epartement de Physique Nucl\'eaire,
Universit\'e Paris-Saclay, F-91191 Gif-sur-Yvette}

\author{David Regnier$^*$} 
\email{david.regnier@cea.fr}
\affiliation{Centre de math\'ematiques et de leurs applications, CNRS, ENS Paris-Saclay, Universit\'e Paris-Saclay, 94235, Cachan cedex, France}
\affiliation{CEA, DAM, DIF, 91297 Arpajon, France}

\author{Jean-Paul Ebran} \email{jean-paul.ebran@cea.fr}
\affiliation{CEA, DAM, DIF, 91297 Arpajon, France}

\author{Antonin Penon} \email{penon@magic-lemp.com}
\affiliation{Magic LEMP, 2 Rue Jean Rostand, 91400 Orsay, France}

\keywords{Artificial Intelligence, Machine Learning, Nuclear Structure, Energy Density Functional, Beyond Mean Field, Active Learning}

\title{Taming nuclear complexity with a committee of multilayer neural networks}

\begin{abstract} 
We demonstrate that a committee of deep neural networks is capable of predicting the ground-state and excited energies of more than 1800 atomic nuclei with an accuracy akin to the one achieved by state-of-the-art nuclear energy density functionals (EDFs) and with significantly less computational cost. 
An active learning strategy is proposed to train this algorithm with a minimal set of 210 nuclei. 
This approach enables future fast studies of the influence of EDFs parametrizations on structure properties over the whole nuclear chart and 
suggests that for the first time a machine learning framework successfully encoded several correlated aspects of nuclear deformation.
\end{abstract}

\maketitle





\paragraph{Introduction}

Today, more than 3000 atomic nuclei have been identified, revealing the wide diversity of nuclear phenomenology (deformation, superfluidity, clustering, halo, \dots ).
Predicting nuclear properties over the whole range of known mass and charge and beyond is therefore a daunting challenge yet essential to unveil new exotic states of matter, foster our quest to the super heavy island of stability and answer the fundamental questions of nucleosynthesis.
Among the different microscopic approaches of nucler structure, only the framework of energy density functionals (EDFs)~\cite{schunck_energy_2019} is currently capable of providing a complete and accurate description of ground- and excited-state properties~\cite{bender_going_2008,niksic_relativistic_2011,egido_state---art_2016}.
Large-scale deployment of nuclear EDFs is however associated with a high computational cost, especially when it is implemented at its multireference level (MR-EDFs), also referred as beyond mean field. 
Such a cost is prohibitive to our understanding of the variations of global nuclear features with different EDFs. Ultimately, fitting an effective interaction at the beyond mean field level is a tremendous task that was only undertaken once~\cite{goriely_first_2009} and yielded a root mean square error (RMS) on the experimentally known masses of 790 keV. As a result most of our beyond mean field calculations are based on EDFs fitted at the mean field level which brings a double counting bias in the predictions.
Attacking the problem from a different angle, Athanassopoulos \textit{et. al.} built a neural network capable of predicting the whole nuclear table of mass with a RMS of 950 keV~\cite{athanassopoulos_nuclear_2006}.
This idea has then been further explored by training neural networks, bayesian networks or gaussian processes to predict the residual between an existing theory and experimental data. It was applied on different models and observables (masses, charge radii and two neutron separation energies) and reduces typically the binding energy RMS to a few  hundreds of keV~\cite{utama_refining_2017,zhang_performance_2017,utama_validating_2018,neufcourt_bayesian_2018,niu_nuclear_2018, akkoyun_artificial_2013,utama_nuclear_2016}.
In all these studies, the quality of the predictions is obtained (i) by the knowledge of an initial model with good performances (typicaly 1-2 MeV RMS on the ground state mass), (ii) by training the artificial intelligence (AI) on  a vast amount of experimental data (especially masses and radii), typically 80\% of one of the Atomic Mass Evaluations (AME)~\cite{wang_ame2016_2017} i.e. more than 1800 nuclei.
This large training set, as well as the fact that these algorithms can only predict one observable, severely restricts the predictive capability of such fast approaches as compared to the EDF approach.
In this letter, we propose a new strategy where an algorithm learns not one observable but several intermediate quantities (potential energy surfaces and inertia) involved in a multireference EDF approach. 
The idea is that while speeding up drastically the calculation of these quantities, the AI will encompass the correlations between several aspects of nuclear deformation.
After a training step, this approach also enables to compute from the AI's predictions multiple low-energy observables such as the ground state and excited energies.

\paragraph{Method}

In this work, nuclear structure properties are tackled within the five-dimensional collective Hamiltonian (5DCH) approach~\cite{kumar_complete_1967,libert_microscopic_1999,delaroche_structure_2010,fu_beyond_2013}.
First, constrained Hartree-Fock-Bogoliubov (HFB) calculations in the space
spanned by both the axial and triaxial quadrupole mass moments capture the static correlations associated 
to quadrupole deformation and pairing. 
The generation of this manifold of states labeled by the quadrupole deformation variables $(\beta,\gamma)$, is by far the most demanding in terms of numerical ressource.
Then, the effect of the quantum-mechanical fluctuations of the order 
parameters $\beta$ and $\gamma$ around the minimal energy region is accounted for through the construction and solving of 
the 5DCH. More precisely, we first compute from the HFB constrained states the corresponding HFB energy surface $E_{\text{HFB}}(\beta,\gamma)$, the $2\times 2$ symmetric matrix $\bm{B}(\beta,\gamma)$ that stands for the vibrational inertia, the rotational inertia $\mathcal{I}_k(\beta,\gamma)$ associated with the three axes $k$ of the intrinsic frame, and a zero-point energy correction $\Delta V(\beta,\gamma)$.
We then build the 5DCH as recalled in Ref.~\cite{supmat} and seek its eigensolutions $g_i(\beta,\gamma,\bm{\Omega})$:
\begin{equation}
\label{eq:collh}
 \left(
 \hat{\mathcal{H}}_{K,rot} + \hat{\mathcal{H}}_{K,vib} + \hat{\mathcal{H}}_{V}
 \right) 
 g_i(\beta,\gamma,\bm{\Omega}) = E_i \, g_i(\beta,\gamma,\bm{\Omega}).
\end{equation}
The collective Hamiltonian contains (i) a kinetic term $\hat{\mathcal{H}}_{K, rot}$ associated with rotation that couples the quadrupolar degrees of freedom to the Euler angles $\bm{\Omega}$, (ii) a vibrational kinetic term $\hat{\mathcal{H}}_{K,vib}$, (iii) a potential term $\hat{\mathcal{H}}_{V}$ that only depends on the quadrupolar deformations.
The eigensolutions of Eq.~\eqref{eq:collh} directly yield the correlated ground state energy as well as the typical rotational and vibrational bands of the excitation spectrum. 

The main idea of this work is to simultaneously teach the eight functions $E_{\text{HFB}}, \Delta V, \bm{B}_{00}, \bm{B}_{01}, \bm{B}_{11}$ and $\mathcal{I}_k\ (k=1,2,3)$ defining the collective Hamiltonian to an AI so that it learns their underlying correlations.
Our AI consists of a committee of multilayer neural networks~\cite{bishop_pattern_2006} that undertake the regression of these functions. 
Each neural network takes as input the number of neutrons $N$ and protons $Z$ and returns the values of the eight functions on a discretized mesh of the deformation space. After a learning stage involving a random initialization of each member, the prediction of the committee is obtained by averaging the outputs of its members. The benefit of using a committee is twofold: (i) reducing the variance of the prediction associated with the random initialization of the members, (ii) providing a simple estimation of this variance which we can leverage in an active learning procedure.

The members of the committee all have the same network architecture. Their input is the number of neutrons and protons encoded in a 600 bits string as detailed in Ref.~\cite{supmat}. 
Note that contrary to Ref.~\cite{niu_nuclear_2018}, we chose an encoding of the inputs that is totally agnostic of any \textit{a priori} knowledge of the physics (\textit{i.e.} shell effects in the vicinity of the valley of stability).
This typically avoids imposing a structure of the inputs based on hard coded magic numbers that may not be relevant in exotic mass and charge regions~\cite{sorlin_nuclear_2008}.
Internally, the neural networks contain five hidden layers defined by the  sequence $600-300-150-100-75$ of their number of neurons.
The first part of the network embeds the information of the nucleus into a neck of 75 neurons only while its second part predicts from this embedding the output functions.
We attempted to fine-tune some hyper-parameters of this architecture such as the number, size and types (dense, convolutions, etc) of the hidden layers with a grid-search approach.
Our results seem quite stable in the neighborhood of the chosen hyperparameters (cf. \cite{supmat}).
%

With this choice of architecture, we perform a supervised training on a set of nuclei for which we know the targeted functions from previous constrained HFB calculations. 
To maximize the quality of the committee while minimizing the number of HFB calculations required for its training we implemented an active learning procedure inspired by Ref.~\cite{yao_speeding_2019}.
It consists of an iterative algorithm which can be summarized by these few steps:
\begin{enumerate}
 \item Sample an initial training set of nuclei and compute their collective functions with constrained HFB.
 \item Train each member of the committee on this set.
 \item Query from the committee a set of additional nuclei that are likely to improve the predictions of the committee if added in the next training step.
 \item Compute the collective functions of these new nuclei with constrained HFB and add them into the training set.
 \item Re-iterate from step 2 up to some stopping criteria.
\end{enumerate}

At step 2, we train independently the neural networks correponding to each member of the committee following a standard technique in machine learning. This procedure, detailed in Refs~\cite{supmat,chollet_deep_2017,abadi_tensorflow:_2015,reddi_convergence_2018,kingma_adam:_2014}, minimizes a training loss while avoiding overfitting the network. 
The training loss consists on a weighted sum of the partial losses $\mathcal{L}_t(N,Z)$ per nucleus ($N,Z$) and per output function $t$.
The partial losses are themselves defined as the squared error between the AI's prediction $t_{\text{AI}}$ and the HFB calculation $t_{\text{HFB}}$ averaged on the deformation space $(\beta,\gamma)\in [0,\beta_+=0.9]\times [0, \frac{\pi}{3}]$ for one nucleus:
\begin{equation}
\mathcal{L}_t(N,Z) = \frac{6}{\pi \beta_+^2}
 \int_{\beta,\gamma}
|t_{\text{AI}}(\beta,\gamma) - t_{\text{HFB}}(\beta,\gamma)|^2   
\text{d}\beta \beta \text{d}\gamma.
\end{equation}

After each training stage, five new nuclei are added to the training set. To select them, we improved the method proposed in Ref.~\cite{yao_speeding_2019} in the following way. 
Each member of the committee makes a prediction for more than 2000 nuclei and we first isolate the 10\% for which the standard deviation between members is the highest.  Then we use a k-means algorithm to detect five clusters among these nuclei and take in each cluster the nucleus for which members predictions differ the most.
To accelerate the training process, we normalized each output function (cf. \cite{supmat}).
The HFB energy is for instance transformed by first removing a deformed liquid drop formula inspired by Ref.~\cite{wang_modification_2010} and then performing a linear scaling so to obtain a zero mean and unity standard deviation on the training set of nuclei.

\paragraph{Results}

We consider a data set of 2100 even even nuclei taken from the AMEDEE database~\cite{hilaire_large-scale_2007} and with the charge and neutron ranges $Z\in[10,120]$ and $N\in[10,260]$.
For these nuclei, we dispose of the eight functions  $E_{\text{HFB}}, \Delta V, \bm{B}_{00}, \bm{B}_{01}, \bm{B}_{11}$ and $\mathcal{I}_k\ (k=1,2,3)$ calculated at 94 deformation points with the Gogny D1S effective interaction~\cite{berger_time-dependent_1991}. For each nucleus, we first interpolated these raw data on a $64\times64$ regular grid with splines of degree two.
We start the active learning of the committee of NN with 2\% of the nuclei sampled randomly from an adapted jittered sampling~\cite{pausinger_optimal_2018}, that ensures a certain uniformity in the $N$ $Z$ plane.
We then run the active learning up to a point where the training set contains 20\% of the 2100 nuclei. At each learning step, we evaluate the quality of the committee's predictions on the $M_{\text{test}}$ test nuclei not present in the training set ($M_{\text{test}} > 80\%$ of the database). To do so we determine for each output function $t$ its associated RMS defined as:
\begin{equation}
 \label{eq:rms}
 \text{RMS}(t) = \left( 
 \frac{1}{M_{\text{test}}} \sum_{i}^{M_{\text{test}}} \mathcal{L}_t(N_i,Z_i)
 \right)^{1/2}
\end{equation}
%
%
\begin{figure}[t]
 \includegraphics[width=0.90\linewidth]{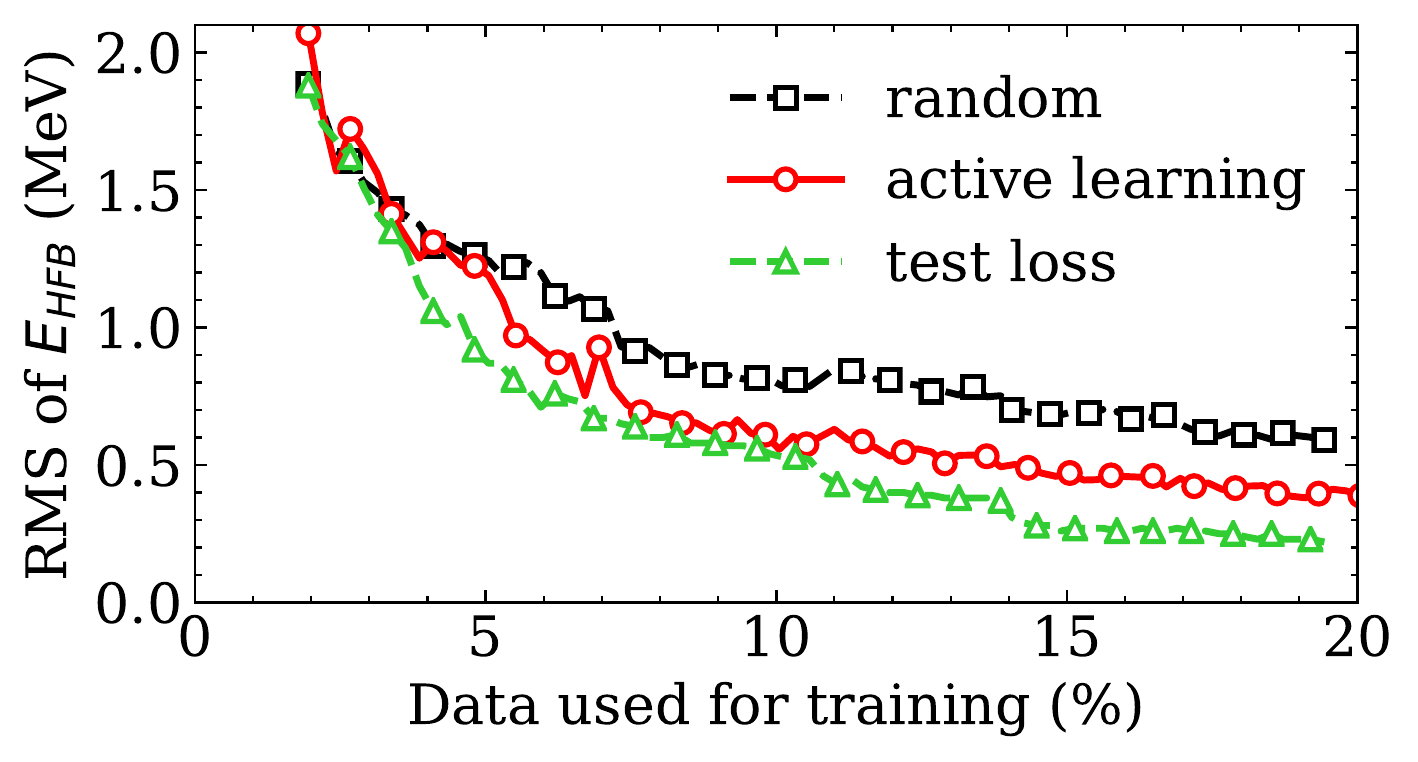}
 \caption{RMS of the HFB energy on the test nuclei as a function of the size of the training set (in percent of the AMEDEE database). We compare results obtained with a training set determined by the active learning (red), by random sampling (black) and by an incremental choice based on the test loss (green).}
 \label{fig:training_rms}
\end{figure}
%
We show in Fig.~\ref{fig:training_rms} the evolution of the RMS associated with the HFB energy as a function of the size of the training set. Starting above 2 MeV, it follows an exponential like decrease to reach less than 400 keV at 20\%.
We compare these results with the ones obtained if we train the same committee of NN on a set of nuclei that is (i) randomly chosen at each step (ii) incremented at each step with the 5 nuclei that maximize the global test RMS (computed on the test set). Note that this last procedure requires the \text{a priori} knowledge of the HFB results for all nuclei. 
The active learning approach outperforms the na\"ive random selection of the training set by roughly 200 keV as soon as more than 5\% of the data set is used for training.
In addition, the training based on the test loss gives even better results in this region. This shows the possibility that a more sophisticated algorithm of active learning could still improve our current results.
Choosing the size of the training set is a tradeoff between the accuracy of the resulting AI and the numerical cost associated with the HFB calculations of the training nuclei.
In Tab.~\ref{tab:ref_rms}, we report the RMS of the eight output functions obtained at four different steps of the active learning.
A striking result is the quality of the committee's prediction already achieved with only 10\% of the total dataset.
The HFB energy, which is a key feature in the determination of the correlated energies, is reproduced within 557 keV over the 1890 nuclei of the test set. In the following we therefore show the results obtained with this 10\% training set.

%
\begin{table}[t!]
 \begin{tabular}{c|cccccccc|c}
 \hline
 Training & $E_{\text{HFB}}$  & $\Delta V$  & $\mathcal{I}_1$ & $\mathcal{I}_2$ & $\mathcal{I}_3$ & $\bm{B}_{00}$ & $\bm{B}_{01}$ & $\bm{B}_{11}$ & $E_{\text{GS}}$ \\ 
 \% 
 & \multicolumn{2}{c}{(keV)}  
 & \multicolumn{3}{c}{($\hbar^2\times$MeV$^{-1}$)} 
 & \multicolumn{3}{c|}{(MeV$^{-1}$)} 
 & (keV) \\ 
 \hline
 5                 & 1190 & 417 & 1.84 & 2.80 & 0.97 & 13.8 & 12.0 & 28.2 & 1325 \\
 \textbf{10}       & \textbf{557} & \textbf{312} & \textbf{1.40} & \textbf{2.25} & \textbf{0.76} & \textbf{11.7} & \textbf{10.2} & \textbf{23.9} & \textbf{716}\\
 15                & 471  & 247 & 1.25 & 2.02 & 0.69 & 10.6 & 9.4  & 21.9 & 655 \\
 20                & 388  & 202 & 1.22 & 1.96 & 0.68 & 10.2 & 9.1  & 21.2 & 518 \\
 \hline
 \end{tabular}
 \caption{RMS obtained on the test set at different stages of the active learning. The first column contains the size of the training set in \% of the AMEDEE database while the others highlight the RMS of the outputs of the committee of NN. The last column contains the RMS associated with the correlated ground state energy $E_{\text{GS}}$ solution of Eq.~\eqref{eq:collh}.}
 \label{tab:ref_rms}
\end{table}

\begin{figure}[t!]
 \includegraphics[height=0.5\textheight]{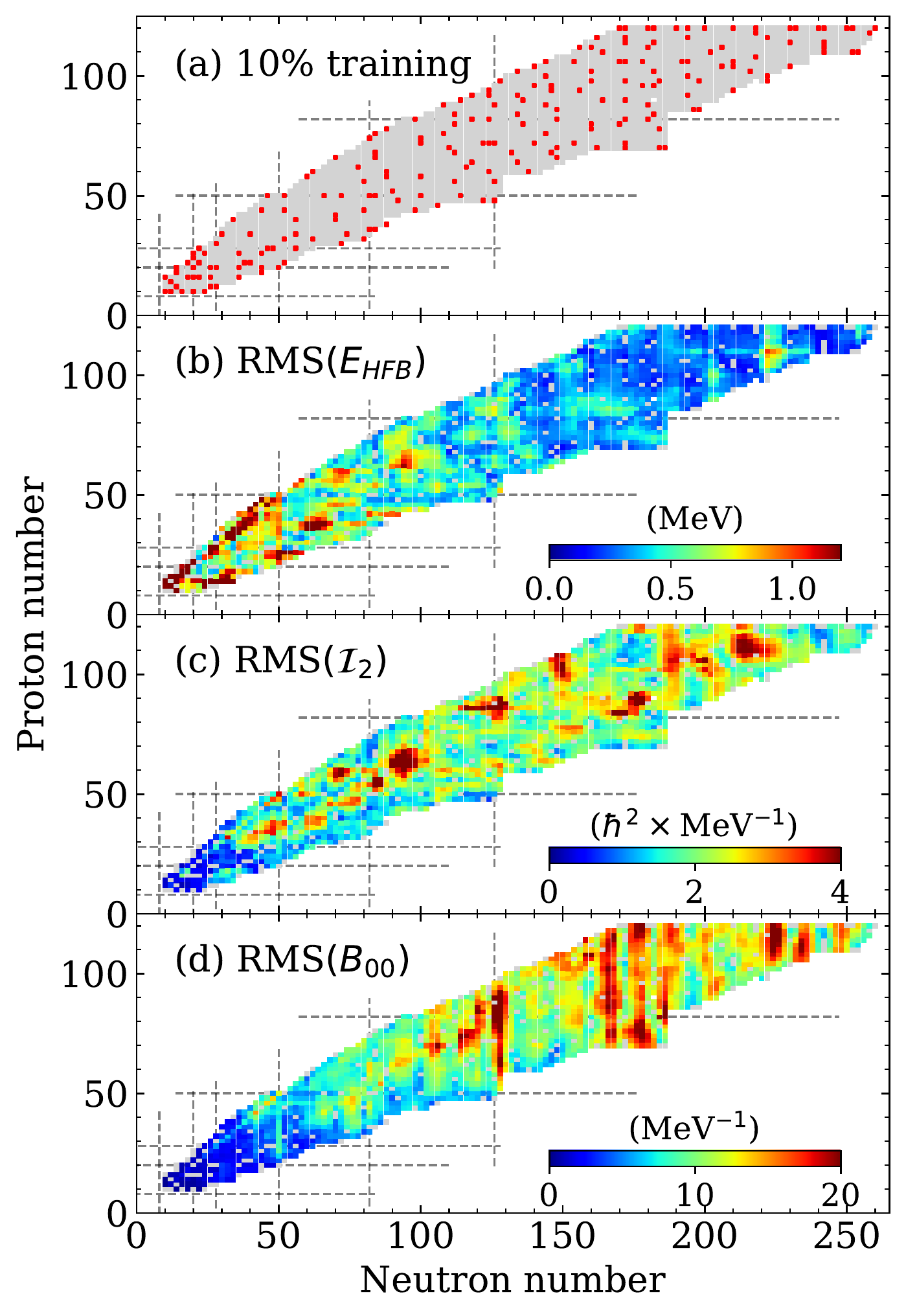}
 \caption{(a) The AMEDEE database nuclei are plotted in grey as a function of $N$ and $Z$. The red squares stand for nuclei included in the 10\% training set obtained by the active learning. The panels (b),(c) and (d) display the resulting AI versus HFB root mean square per nucleus ($\mathcal{L}_t^{1/2}(N,Z)$) for the three outputs $E_{\text{HFB}}$, $\mathcal{I}_2$ and $\bm{B}_{00}$ respectively.}
 \label{fig:rms_chart}
\end{figure}

The quality of the committee's prediction varies with $N$ and $Z$.
To assess what parts of the nuclear chart are correctly grasped by the committee of neural networks, we emphasize in Fig.~\ref{fig:rms_chart} the individual RMS per nucleus $\mathcal{L}_t^{1/2}(N,Z)$ for three different kinds of outputs. 
For the HFB energy, the AI captures very well the vast majority of heavy nuclei but struggles in the medium and light sectors ($N<50$). A specially high RMS is found close to the $N=Z$ line where HFB calculations are known to predict a strong energy cusp.
The difficulty to reproduce the HFB energies in the light sector with neural networks was already encountered in Ref.~\cite{niu_nuclear_2018} and is related to the sharp variations present in this region. The panel (a) of Fig.~\ref{fig:rms_chart} shows that the active learning procedure automatically densified the training set in this region to mitigate this difficulty.
Concerning the vibrational and rotational inertia the error of the AI globally increases with the mass and some of the error peaks can be identified close to shell closures, e.g. for the vibrational inertia close to the neutron number $N=80$.
%
%
\begin{figure}[t!]
 \includegraphics[height=0.5\textheight]{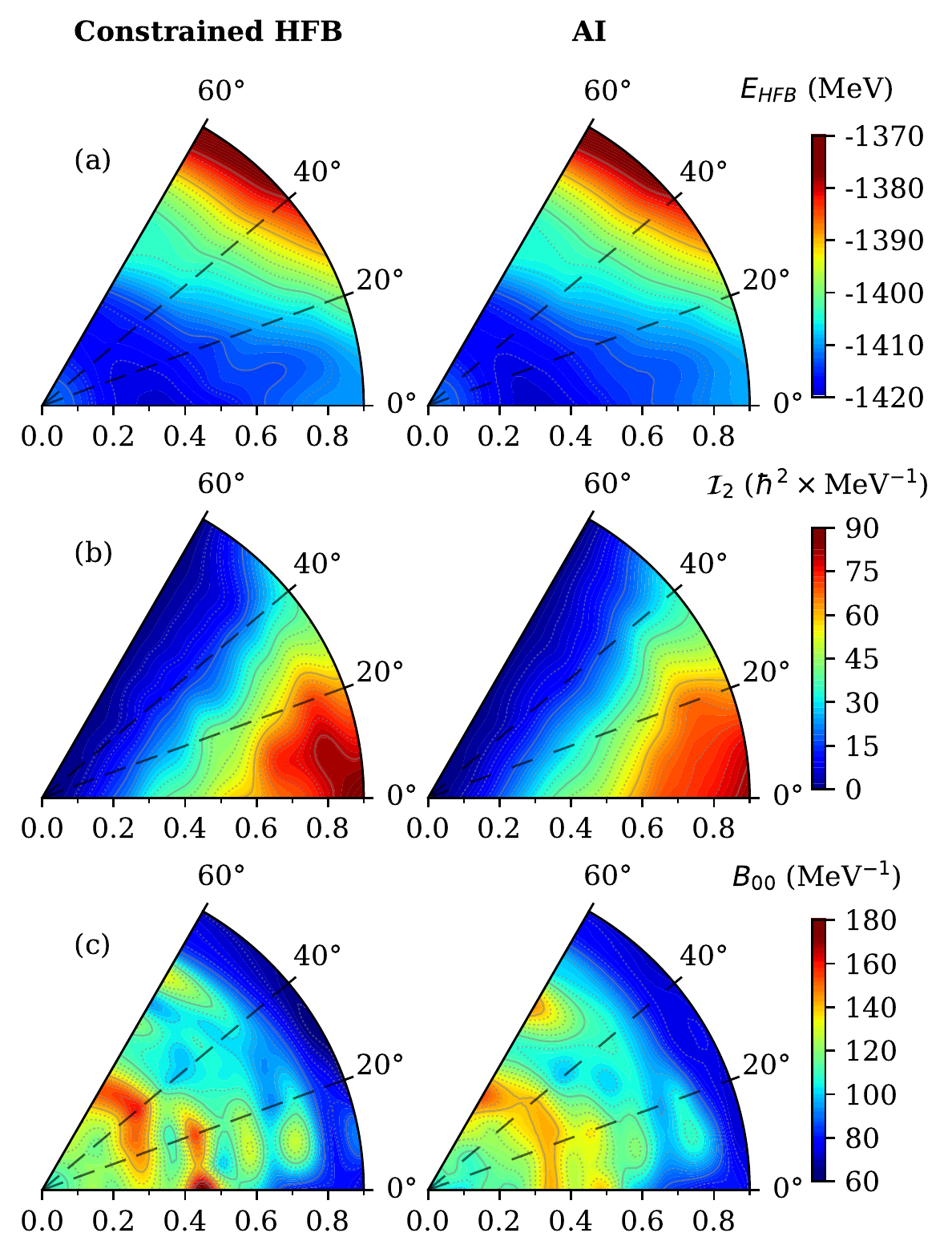}
 \caption{HFB energy (a), rotational inertia along the principal axis (b) and vibrational inertia related to elongation (c) resulting from both constrained HFB calculations and the AI. These functions are plotted for \nuc in the standard polar representation where $\beta$ is the radial coordinate and 
 $\gamma$ the polar angle.}
 \label{fig:pes_compare}
\end{figure}
%
We compare in Fig.~\ref{fig:pes_compare} the AI and reference HFB predictions for three targeted functions for \nuc. 
We choose this nucleus because (i) its excitation spectrum is known experimentally and (ii) its partial root mean square $\text{RMS}(E_{\text{HFB}})$ is $409$ keV which lies just above the median of this quantity over the test set. It is therefore representative of how the committee of NN performs for most of the test nuclei.
Once again, the overall topology of $E_{\text{HFB}}$, $\mathcal{I}_2$ and $\mathcal{B}_{00}$ are very well grasped by the committee despite the fact that the closest nucleus in the training set is the $^{180}$W which has 4 additional neutrons and 2 protons less.

\begin{figure}[!ht]
 \includegraphics[width=0.90\linewidth]{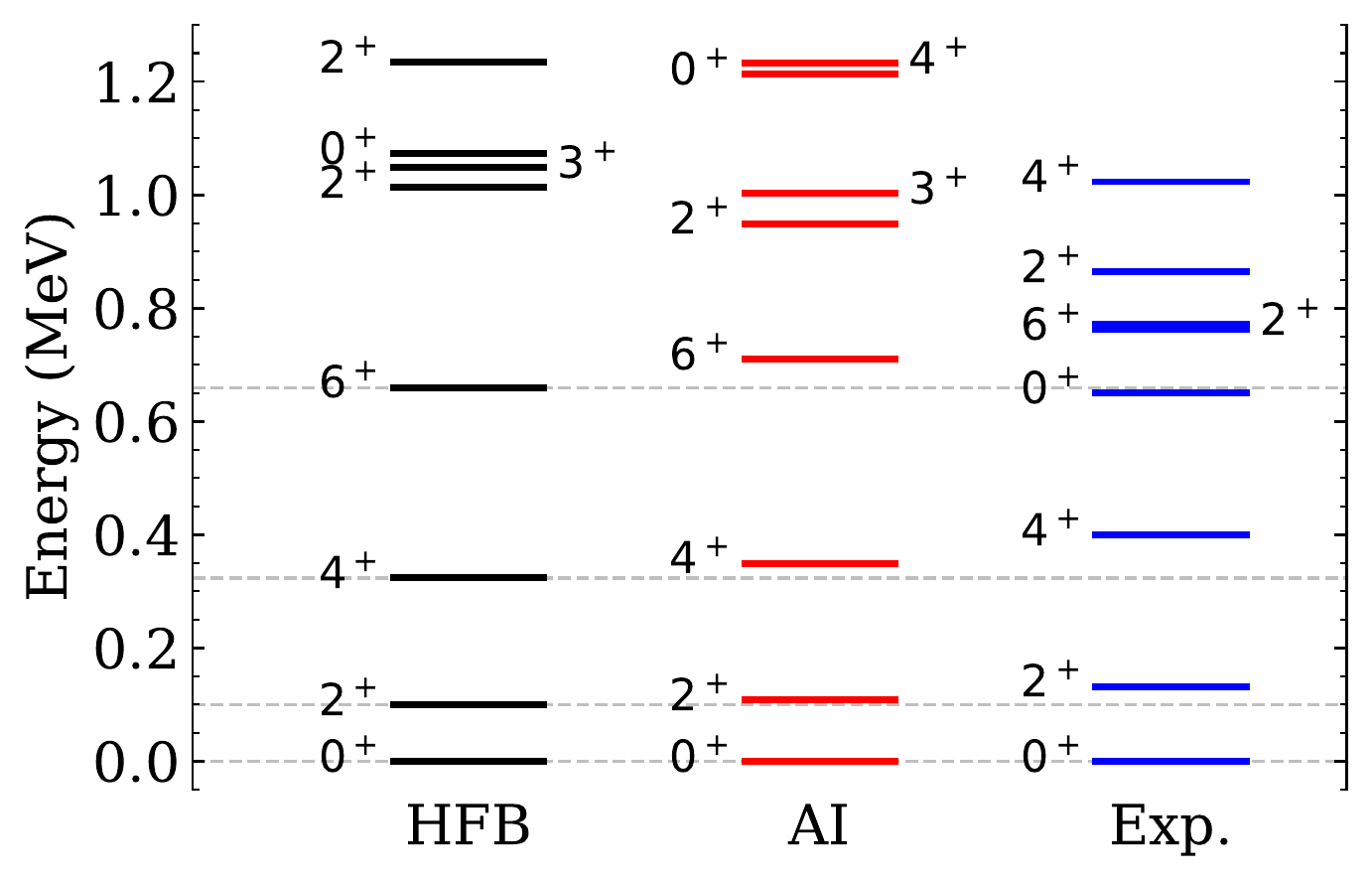}
 \caption{Excitation spectrum of \nuc obtained from both the AI (AI) and the constrained HFB calculations (HFB). The experimental spectrum (Exp.), taken from the ENSDF database~\cite{noauthor_evaluated_nodate}, is also displayed.}
 \label{fig:spectrum}
\end{figure}

Finally we focus on the correlated ground state energy and excitation spectra obtained from Eq.~\eqref{eq:collh}.
To solve the eigenproblem, we discretized the Euler angle space on the basis of Wigner rotational wave functions as in Ref.~\cite{kumar_complete_1967} whereas the deformation space is discretized on a finite element basis implemented with the FELIX-2.0 library~\cite{regnier_felix-2.0:_2018}. 
We perform the comparison between the AI and the HFB reference for the 1666 nuclei of the test whose minimum of the potential energy lies below $\beta=0.8$. This simple filter removes the super-heavy nuclei with an open fission channel in our deformation space, for which the ground state would be spuriously predicted at too high deformations. 
As reported in Tab.~\ref{tab:ref_rms}, we obtain a RMS of 716 keV for the ground state energy.
Although a direct comparison is not sound, note that this number has the same order of magnitude than the predictions to experiments root mean square obtained with state-of-the-art nuclear EDFs (500-800 KeV for the Skyrme HFB mass models~\cite{chamel_further_2008,goriely_skyrme-hartree-fock-bogoliubov_2009}).
Fig.~\ref{fig:spectrum} displays the excitation spectra of \nuc obtained from both the HFB and committee predictions and give for the sake of completeness the experimental values taken from the ENSDF database~\cite{noauthor_evaluated_nodate}. 
The rotational band predicted by the committee of NN impressively matches the HFB data with only a 8\% deviation for the first $2^+,4^+$ and $6^+$ states. Finally, the first excited $0^+$ level is reproduced within 13\% despite the complexity of the vibrational inertia topology.

\paragraph{Conclusions}

In this work we built, for the first time, a machine learning framework capable of estimating the low energy structure of all nuclei from a given EDF.
Stunning performances are achieved, viz. a RMS of 716 KeV on the correlated ground state energy with respect to the 
the MR-EDF calculation for a training on only $\sim$ 200 nuclei.
Further improvements seems at reach, e.g. by (i) refining the selection of the training set of nuclei (cf. Fig.~\ref{fig:training_rms}), (ii) exploring more involved active learning techniques such as negative correlation learning~\cite{liu_ensemble_1999}, or more sophisticated kinds of neural networks.
This fast framework opens the opportunity to quickly test the impact of new parametrizations of EDFs in the context of astrophysics and super-heavy production.
On top of this, it paves the way toward fitting new EDFs at the multireference level and with multiple observables (ground state masses, radii, and spectroscopic features).
Finally, the success of this approach is a first proof of principle that an committee of NN is able to encode several correlated aspects of nuclear deformation.
The neural networks involved likely possess a satisfying non-trivial internal representation of the physics of the system. 
Studying this representation may unveil new physical concepts grasped during the active learning.

\begin{acknowledgments}
The authors would like to thank J. Libert, S. Hilaire, G. Hupin and E. Khan for fruitful discussions on the collective Hamiltonian and the nuclear mass tables, as well as N. Makaroff for his suggestion on activation functions.
{$^*$ D.~Regnier and R.~Lasseri contributed equally to this work.}
\end{acknowledgments}


%

\end{document}